\documentclass[aps,prl,reprint,a4paper,superscriptaddress,citeautoscript,floatfix]{revtex4-2}

\usepackage{amsmath,amsfonts,amssymb}	
\usepackage{newtxtext,newtxmath}		
\usepackage{mathrsfs}					

\usepackage[pdftex]{graphicx}
\usepackage{microtype}
\usepackage{bm}

\usepackage[svgnames]{xcolor}
\usepackage[colorlinks=True,linkcolor=DarkRed,citecolor=ForestGreen,urlcolor=MediumBlue,
	pdfstartview=FitH,bookmarks=False,pdfpagemode=UseNone]{hyperref}

\def\kF{k_{\mathrm{F}}}

\begin{document}

\title{Localization Transition for Interacting Quantum Particles in Colored-Noise Disorder}

\author{Giacomo Morpurgo}
\affiliation{Department of Quantum Matter Physics, University of Geneva, 1205 Geneva, Switzerland}
\author{Laurent Sanchez-Palencia}
\affiliation{CPHT, CNRS, Ecole Polytechnique, IP Paris, F-91128 Palaiseau, France}
\author{Thierry Giamarchi}
\affiliation{Department of Quantum Matter Physics, University of Geneva, 1205 Geneva, Switzerland}

\date{\today}

\begin{abstract}

We investigate the localization transition of interacting particles in a one-dimensional correlated disorder system. The disorder which we investigate allows for vanishing backwards scattering processes. We derive by two renormalization group procedures its phase diagram and predict that the localization transition point is shifted from finite attractive interaction to the non-interacting point. We finally show numerically that the scaling of the localization length with the disorder strength deviates from the usual scaling of a localized phase.
\end{abstract}

\maketitle

The discovery of Anderson localization (AL) \cite{Anderson_1958_localization} has shown that disorder can profoundly change the electronic properties of noninteracting materials, leading to extensive studies of the effects of disorder in quantum systems. Its effect is even more dramatic in low dimensions where an infinitesimal amount of disorder is enough to completely change the nature of electronic wavefunctions and lead to localization. In this context, a major challenge is to understand the combined effects of disorder and interactions \cite{Altshuler_Lee_1980_disorder_interaction_2d, Finkelstein_1984_localization_interaction, Review_Lee_Ramakrishnan_1985_disorder}. Again, the most dramatic competition occurs in low dimensional systems, since both the effects of interactions \cite{Book_Giamarchi_2004_1D} and disorder are maximal there. This induces many novel phenomena, including the emergence of Bose glass phases \cite{Giamarchi_Schulz_1988_1D_Anderson_localization, Fisher_Fisher_1989_Bose_Glass} and many-body localization \cite{Review_Abanin_Serbyn_2019_MBL, Luitz_Alet_2015_MBL_mobility_edge, Imbrie_2016_MBL_proof}.

Most of the studies mentioned above deal with spatially uncorrelated (white-noise) disorder.
However, correlations in disorder can strongly affect the behaviour of various systems, ranging from crystallography to superconductivity and Anderson localization~\cite{Review_Keen_Goodwin_2015_correlated_disorder_crystallography,Neverov_Croitoru_2022_correlated_disorder_enhance_superconductivity,Review_Sanchez_Palencia_Lewenstein_2010_disorder_cold_atoms}. The realization of other types of disorder-like potentials in quantum-simulation platforms such as ultracold-atom and cavity polariton systems prompted a large body of theoretical and experimental work. For instance, quasiperiodic models have been extensively studied for quantum systems with or without interactions~\cite{Roati_Inguscio_2008_Anderson_localization_quasiperiodic,Tanese_Akkermans_2014_Polaritons_in_Quasiperiodic,Aubry_Andre_1980_quasiperiodic_anderson_localization,Vidal_Giamarchi_1999_quasiperiodic_potential_short,Vidal_Giamarchi_2001_quasiperiodic_potential_long,Roux_Giamarchi_2008_Quasiperiodic_Bose_Hubbard_localization,D_Errico_Modugno_2014_bose_glass,Gori_Roux_2016_temperature_disorder_bosons_1D,Yao_Sanchez_Palencia_2019_quasiperiodic_1d_equal_potential,Yao_Sanchez_Palencia_2020_quasiperiodic_equal_potential_bose_glass,Schreiber_Bloch_2015_MBL_cold_atoms}. Another interesting case is that of speckle potentials as realized in cold atomic systems, which strongly alter AL~\cite{Sanchez_Palencia_Aspect_2007_Localisation_Speckle_Bosons,Billy_Aspect_2008_Anderson_localization_speckle,Kuhn_Muller_2005_localization_2d_atoms_optical_lattices,Kuhn_Muller_2007_transport_speckle_potential,Piraud_Sanchez_Palencia_2012_3D_anisotropic_disorder}. In 1D, suppression of backscattering by speckle correlations leads to an effective suppression of AL and the appearance of a pseudo-mobility edge, where the localization length changes by several orders of magnitude~\cite{Lugan_Miniatura_2009_Anderson_localization_speckle}. This can be exploited to control AL by correlation engineering in speckle potentials~\cite{Piraud_Sanchez_Palencia_2012_tailored_disorder,Piraud_Sanchez_Palencia_2013_Anderson_localization_tailored_disorder_1d_speckle}. Studies have also been led to investigate AL of collective excitations in weakly interacting 1D bosons subjected to speckle disorder within Bogoliubov formalism~\cite{Gurarie_Chalker_2002_bosons_disorder,Gurarie_Chalker_2003_bosons_disorder_long,Lugan_Sanchez_Palencia_2007_Anderson_localization_Bogolyubov,Lugan_Sanchez_Palencia_2011_Anderson_localization_Bogolyubov_correlated_disorder}. So far, however, the effect of speckle-like correlations on the localization of quantum particles in the strongly interacting regime (bosons or fermions) remains unknown.

In this Letter, we show that for an interacting 1D quantum system with spatially correlated (colored noise) disorder of the speckle type, in the case where backscattering is vanishingly allowed, the 
localization-delocalization transition is drastically modified compared to standard white noise disorder.
Using universal Tomonaga-Luttinger liquid (TLL) theory, we treat bosons and fermions on equal footing.
Most importantly, renormalization group (RG) analysis shows that the critical point is shifted from the Luttinger parameter $K^*=3/2$ (white noise) to $K^*=1$ (speckle-like). This result is confirmed using direct RG analysis of the microscopic interacting Fermi model.
We also find that the localization length presents an unusual scaling with the disorder strength. Our findings show that special type of disorder correlations can strongly alter the critical properties of interacting quantum systems. They directly apply to speckle potentials as well as other kinds of colored noise disorder, which can be implemented using digital mirror devices (DMD) in cold atomic systems. 

Using the bosonized representation \cite{Haldane_1981_Bosonization, Book_Giamarchi_2004_1D}, which describes well the low-energy properties of the system, the Hamiltonian reads:
\begin{align} 
    H &= H_{0} + H_{W} \label{eq:hamtotal} \\ 
    H_{0} &= \frac{1}{2\pi}\int dx \, u\left[K\big(\nabla\theta(x,\tau)\big)^2 + \frac{1}{K}\big(\nabla\phi(x,\tau)\big)^2\right]\\
    H_{W} &= \frac{1}{\pi \alpha}\int dx \, W(x)\cos\big(2\phi(x) -2\kF x\big)
\end{align}
where $\phi$ and $\theta$ are two bosonic fields with the commutation relation $\left[\frac{1}{\pi}\nabla\phi(x),\theta(x')\right] = -i\delta(x -x')$, $u$ is the speed of sound, and $K$ is the dimensionless (interaction-dependent) Luttinger parameter. This representation applies to fermionic, bosonic or spin systems both in the continuum or a lattice. The quantity $\kF = \pi\rho_0$, with $\rho_0$ the average density, is the Fermi wave vector and $\alpha$ is a ultraviolet cutoff of the order of the lattice spacing for a lattice model. For fermionic systems, $K = 1$ corresponds to free fermions while $K > 1$ (resp.~$K < 1$) corresponds to attractive (resp.~repulsive) interactions. For bosonic systems with contact repulsive interactions, $K \in [1,\infty[$, with $K=1$ corresponding to infinite repulsion (Tonks limit) and $K\to+\infty$ to free bosons~\cite{Book_Giamarchi_2004_1D, Review_Cazalilla_Giamarchi_2011_1D_bosons}.

\begin{figure}
\includegraphics[width=0.95\columnwidth]{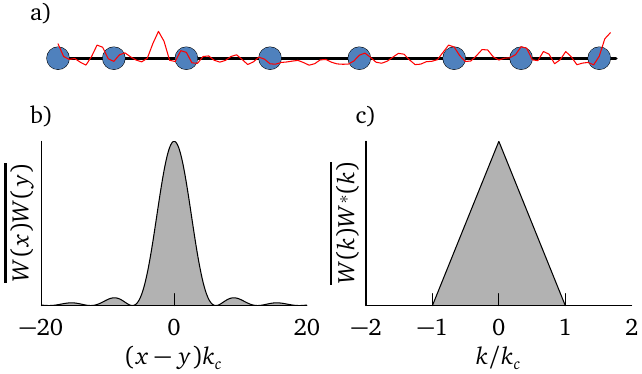}
\caption{\label{fig:model}
Sketch of a colored-noise disorder as considered in this work. (a)~Realization of a 1D speckle potential (red line) with quantum particles (blue disks). (b)~Two-point correlation function of a speckle potential in real space. (c)~Two-point correlation function of a speckle potential in momentum space. The latter has a triangular form with a high-momentum cut-off at $|k|=k_c$.
}
\end{figure}

The term $H_W$ represents the backscattering of particles with momentum close to $\kF$ (scattering with transfer momentum $2\kF$) from a disorder potential $W(x)$ as shown in Fig.~\ref{fig:model}(a).
In 1D, the forward and backward scatterings can be decoupled \cite{Giamarchi_Schulz_1988_1D_Anderson_localization}. Since only the backscattering affects the current, we neglect the forward component of the disorder in this study (See the Supplementary Material (SM)~\cite{supp_mat} for more details on the forward scattering).

We consider a colored noise disorder that can potentially make the backscattering vanish. One example is the speckle disorder~(SD)~\cite{Book_Dainty_2013_Speckle,Book_Goodman_2015_statistical_optics}, which has been instrumental in observing single-particle AL in cold atomic gases~\cite{Billy_Aspect_2008_Anderson_localization_speckle,Kondov_DeMarco_2011_3D_Anderson_localization_speckle,Jendrzejewski_Bouyer_2012_3D_anderson_localisazion_cold_atoms}. It stems from the square of a random, Gaussian correlated, complex field, so that it is non-Gaussian and non-symmetric. Its two-point spatial correlation function is a squared sinc function, see Fig.~\ref{fig:model}(b). The speckle disorder has finite momentum support such that the second moment of the correlations of the potential vanishes above a momentum cut-off $k_c$, see Fig.~\ref{fig:model}(c). It can be either repulsive (blue detuned, BSD) or attractive (red-detuned, RSD). In order to disentangle some of the effects due to the colored noise from the existence of odd moments for BSD and RSD due to their non-Gaussian character, we also study a Gaussian, colored disorder (GCD) with the same correlation function. In both cases, the second moment reads
\begin{equation} \label{eq:disordercorr}
    \overline{W_k W_{-k'}} = W_0^2\left(1 - {|k|}/{k_c}\right)\theta(k_c - |k|)\, \Omega\, \delta_{k,k'}
\end{equation}
where the overbar denotes the average over disorder realizations, $W_0$ is the disorder intensity, $\Omega$ the volume of the system, and $\delta_{k,k'}$ the Kronecker delta. Note that the factor $\Omega$ comes from the use of discrete values of $k$, and the spatial correlations do not depend on the volume of the system. More details can be found in the SM~\cite{supp_mat}.

To consider the combined effects of disorder and interactions, we treat the disorder term in Eq.~(\ref{eq:hamtotal}) using a perturbative RG procedure. Working along the lines of Ref.~\cite{Vidal_Giamarchi_2001_quasiperiodic_potential_long}, we integrate the short-distance properties by increasing the cutoff $\alpha(l) = \alpha e^l$. This is equivalent to integrate the momenta in a shell around $2\kF$ with width $1/\alpha(l)$. It yields the RG equations~\cite{supp_mat}: 
\begin{align}
    \label{eq:RG_bosonization_K_complete}
    \frac{\partial K}{\partial l} &= -\frac{K^2y^2}{2}\frac{1}{\Omega}\sum_k \left(1-\frac{|k|}{k_c}\right)\theta(k_c-|k|) \\
    &\quad \quad\times\left[J_0\big((k+2\kF)\alpha(l)\big) + J_0\big((k-2\kF)\alpha(l)\big)\right]\nonumber\\
    \label{eq:RG_bosonization_K_complete_bis}
    \frac{\partial y^2}{\partial l} &= (4-2K)y^2,
\end{align}
where $y = \frac{\alpha W_0}{u}$ and $J_0$ is the Bessel function. The latter acts as a ``window'' filtering the modes far from the Fermi wave vector by a value of order $1/\alpha(l)$. At this order, the RG equations depend only on the second moment of the disorder correlations and are thus identical for SD and GCD. The appearance of the Bessel functions is due to our choice of a hard cutoff in real space and, similarly as in Ref.~\cite{Vidal_Giamarchi_2001_quasiperiodic_potential_long}, we replace them by a window $J_0(q) = \theta(1-|q|)$. Intuitively, the RG procedure amounts to making the windows centered at $\pm2\kF$ smaller and smaller, thus capturing only the physics which occur there at low-energy (i.e.~at $\sim \pm2\kF$). 

Three cases of interest arise depending on the value of $k_c$ compared to the Fermi level.
\begin{figure}
\includegraphics[width=0.9\columnwidth]{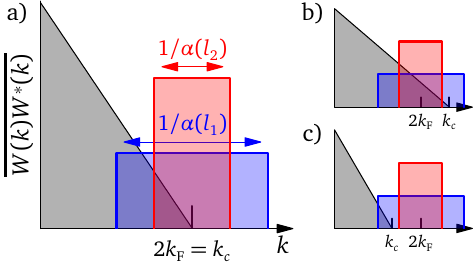}
\caption{\label{fig:RG_procedure} Bosonized RG procedure, as $\alpha(l)$ increases ($l_1<l_2$), the windows around $2\kF$ shrink, capturing only the low-energy physics. (a) Case $k_c = 2\kF$. The weight of the disorder in this window becomes vanishingly small as $\alpha$ increases. (b) Case $k_c > 2\kF$. As $\alpha$ increases, there is always a finite weight of the disorder in the shrinking window. We recover the physics of a uncorrelated Gaussian disorder. (c) Case $k_c < 2\kF$. There is a certain $l^*$ after which there is not any weight of the disorder inside the shrinking window. We recover the physics of non-disordered systems.
}
\end{figure}
For $k_c < 2\kF$, there is a scale $l^*$, such that $\alpha(l>l^*) > 1/(2\kF-k_c)$ and the window does not contain any disorder anymore, owing to the finite support of the latter, see Fig.~\ref{fig:RG_procedure}(c). Hence, no backscattering occurs at this order in the RG, which implies suppression of localization. Note that higher order perturbation terms in the disorder may induce backscattering~\cite{Lugan_Miniatura_2009_Anderson_localization_speckle}. However, for the case of weak disorder that we consider here, such terms corresponding to a higher power of the disorder would be extremely small and lead potentially only to a huge localization length.

For $k_c > 2\kF$, backscattering is always present at all scales. For $\alpha(l) \gg 1/(k_c-2\kF)$, the second moment of the disorder is almost constant within the window, see Fig.~\ref{fig:RG_procedure}(b), and we recover the same RG equations as for an uncorrelated Gaussian disorder \cite{Giamarchi_Schulz_1988_1D_Anderson_localization,Vidal_Giamarchi_2001_quasiperiodic_potential_long}. In that case, the momentum cutoff in the spectrum of the disorder is irrelevant and we find a localization-delocalization transition (for weak disorder) at the usual critical critical point $K^* = 3/2$. 

\begin{figure}
\includegraphics[width=1\columnwidth]{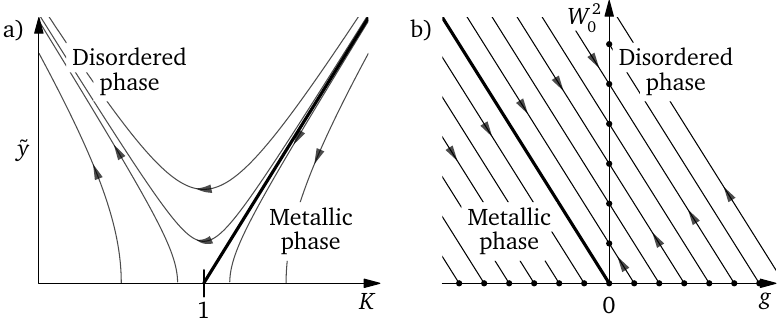}
\caption{\label{fig:phase_diagram}
Phase diagrams. (a)~Sketch of the phase diagram versus Luttinger parameter and renormalized disorder strength, with flow lines from the bosonized RG procedure for $k_c=2\kF$. The separatrix (bold line) separates the disordered (localized) and metallic (delocalized) phases, with the critical point at $K^*$ in the weak disorder limit. (b)~Sketch of the phase diagram versus interaction and disorder strengths, with flow lines from the diagrammatic RG procedure around the non-interacting Fermi point. Here, the flow is made of straight lines and we have a line of fixed point in the absence of interactions ($g=0$).
}
\end{figure}

The most interesting case, and the central point of our study, corresponds to $k_c = 2\kF$ for which backscattering is vanishingly allowed at all scales, see Fig.~\ref{fig:RG_procedure}(a). In this case, as we progress in the RG, the window shrinking around $2\kF$ always contains disorder, but with a smaller and smaller weight. The linear decrease of the spectral weight $\overline{W_k W_{-k'}}$ implies that the sum over $k$ in Eq.~(\ref{eq:RG_bosonization_K_complete}) scales quadratically with $\alpha(l)$, which yields ${\partial K}/{\partial l} \propto -{y^2}/{\alpha^2(l)}$. Introducing $\tilde{y}=y/\alpha(l)$, we then find the RG equations
\begin{align}
     \frac{\partial K}{\partial l} & = -\frac{K^2}{4\pi k_c}\tilde{y}^2 \label{eq:RG_bosonization_K} \\
     \frac{\partial \tilde{y}}{\partial l} & = (1 - K )\tilde{y}. \label{eq:RG_bosonization_y}
\end{align}
The corresponding RG flow is shown in Fig.~\ref{fig:phase_diagram}(a).
It shows that in this case, the critical point is at $K^*=1$, instead of the value $K^*=3/2$ for white-noise disorder. Hence, for $K<1$, any arbitrary weak disorder implies localization (disordered phase). Instead, for $K>1$, we find a localization transition where a finite amount of disorder is necessary to localize while too weak disorder implies delocalization (metallic phase). The shift of the critical point implies that a colored noise having backscattering vanishing linearly at $2\kF$ can dominate only for significantly less attractive interactions for fermions (resp.~more repulsive interactions for bosons) than for standard white noise disorder. This remarkable shift of the transition point is consistent with the intuitive fact that the backscattering at exactly $2\kF$ would be zero for such a disorder. Nevertheless, due to backscattering at finite momenta around $2\kF$, interactions restore localization at $2\kF$. Note that the value of $K^*$ directly relies on the scaling of the disorder correlation functions at the cut-off $k_c$. For correlations behaving as $\left(1-\frac{|k|}{k_c}\right)^\nu$, following the same steps, we obtain a transition point at $K^* = \frac{3-\nu}{2}$. We thus recover the two limit cases~: For $\nu = 0$ (white noise disorder or colored noise disorder with $2\kF<k_c)$, $K^*=3/2$; For $\nu = 1$ (colored noise disorder with $2\kF=k_c$, $K^*=1$. A colored noise disorder with $0<\nu<1$ may be realized using a mask with varying transmission for SD~\cite{Piraud_Sanchez_Palencia_2012_tailored_disorder,Piraud_Sanchez_Palencia_2013_Anderson_localization_tailored_disorder_1d_speckle} or directly engineered using DMD in cold atom experiments.

The linear vanishing of the spectral weight of the disorder for standard SD has the advantage to bring back the transition around the non-interacting point for fermions ($K=1$). This allows us to perform a perturbative RG analysis directly on the microscopic model. Moreover, it avoids the necessity, in the bosonization procedure, to carefully distinguish, around the non-interacting point, between elastic and inelastic processes~\cite{Giamarchi_Schulz_1988_1D_Anderson_localization}, which is something highly non-trivial for the correlated disorder discussed here. To proceed, we consider the Fermi Hamiltonian
\begin{multline}
\label{eq:Hamiltonian_language_fermions}
    H = \sum_r\sum_{k}v_\text{F}(\varepsilon_rk-\kF)c^\dagger_{r,k}c_{r,k}\\ + \frac{g}{2\Omega}\sum_{r}\sum_{k,k',q}c^\dagger_{r,k+q}c^\dagger_{-r,k'-q}c_{-r,k'}c_{r,k}\\
    +\frac{1}{\Omega}\sum_{k, q\sim 0} \left[W_{q-2\kF}c^\dagger_{R,k+q}c_{L,k} + W_{q+2\kF}c^\dagger_{L,k+q}c_{R,k}\right],
\end{multline}
where the kinetic term is linearized around the Fermi momentum, $v_\text{F}$ is the Fermi velocity, $g$ the strength of the interactions, $\varepsilon_r = \pm 1$ for right/left movers and $W_{q-2\kF}$ is defined as before in Eq.~(\ref{eq:disordercorr}). We impose an ultraviolet cutoff in momentum space $\Lambda \sim \frac{1}{\alpha}$, equivalent to the one used in bosonization. We expand up to second order in interactions $g$ and disorder lines $W_0^2$, and look for the 2nd-order diagrams which renormalize the 1st-order interaction and backscattering diagrams. In both cases, there is a single $g W_0^2$ diagram to be computed. We find that both 2nd-order diagrams diverge logarithmically with $\Lambda$. More details on the procedure~\cite{Review_Solyom_1979_Fermi_gas_1d_conductor} can be found in the SM~\cite{supp_mat}. Upon varying the cutoff as $\Lambda(l) = \Lambda e^{-l}$, we find the RG equations
\begin{align}
     \frac{\partial g}{\partial l}& = -\frac{gW_0^2}{2\pi k_c v_\text{F}^2} \\
     \frac{\partial W_0^2}{\partial l}& = \frac{gW_0^2}{2\pi v_\text{F}}.
\end{align}
These equations describe a flow that follows straight lines in the $g-W_0^2$ parameter space, see Fig.~\ref{fig:phase_diagram}(b). This flow confirms the predictions of the bosonization approach. For attractive interactions, $g < 0$, the flow reduces the disorder strength. There is a separatrix between the delocalized (metallic) phase where the disorder fully vanishes and the localized (disordered) phase where it grows up a finite value through RG. Remarkably, we find that the non-interacting line is a line of stable fixed points where the amplitude of the disorder remaining constant under the RG flow.

Besides the phase diagrams, Fig.~\ref{fig:phase_diagram}, computing physical properties, such as the nature of the metallic and disordered phases and transport properties, is challenging. For the phase where the disorder is relevant, higher order disorder correlations must in principle be taken into account, which we leave for future studies. For the metallic phase, one recovers in principle TLL behavior, characterized by power-law decaying correlation functions, and dominant superconducting/superfluid quasi-long range order.  However, this is justified for a disorder with strictly no Fourier component beyond $2\kF$. Higher terms in the disorder, for instance combining the backward scattering with one forward scattering may generate such Fourier components at order $W_0^4$, restoring a strict critical point at $K^*=3/2$. Nevertheless, for $1 < K < 3/2$ the localization length would be extremely large and, far from the critical point, of the order of $\xi \sim(1/D)^{2/(3-2K)}$ with $D =\overline{W(x)W(x)}$. Below such a length, the system would be fully controlled by the colored part of the disorder with its own ``localization''-delocalization transition at $K^*=1$.

\begin{figure}
\includegraphics[width=0.95\columnwidth]{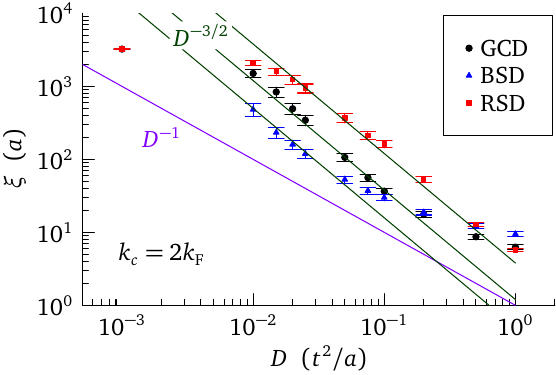}
\caption{\label{fig:localization_length} 
Localization length $\xi$ versus disorder strength $D$ of the eigenstates of a non-interacting system at momentum $\kF = k_c/2$  for GCD (black circles), BSD (blue triangles), and RSD (red squares). The error bars indicate the standard deviation from disorder averaging (1000 realizations). Guides to the eye of $D^{-3/2}$ behavior to the three sets of data are shown as solid green lines as well as the scaling $D^{-1}$ expected for white-noise disorder as a solid purple line for reference. The available range for $\xi$ is limited on one side by the system size (L/a = 10000) and on the other side by the reaching of the strong disorder regime, where usual scalings break down.
}
\end{figure}

To further analyze the consequences of the presence or not of higher moments in the disorder, we may again take advantage of the fact that the non-interacting line is a fixed line by RG in the disordered phase, and consider non-interacting spinless fermions with the tight-binding Hamiltonian
\begin{equation} \label{eq:hamtight}
    H = -t\sum_{\langle i,j\rangle}c^\dagger_ic^{\phantom{\dagger}}_j + \textrm{ H.c } + \sum_i W_i c^\dagger_ic^{\phantom{\dagger}}_i,
\end{equation}
where $t$ is the hopping amplitude, $W_i$ is the disorder potential at site $i$, and $\Omega = Na$, with $a$ the lattice spacing, and $N$ the number of sites. We solve the Hamiltonian in Eq.~(\ref{eq:hamtight}) by exact numerical diagonalization, which therefore contains all scattering orders and includes all forward and backward scattering processes. We use a system of $10000$ sites for all strengths of disorder. To disentangle the roles of the non-Gaussian character of speckles, their non-symmetric property, and of the existence of a spectral cut-off, we consider BSD, RSD, and GCD. In order to compare to uncorrelated Gaussian disorder (white noise), it is convenient in this section to characterize the disorder strength by its zero-distance correlations $D = \overline{W(x)W(x)}= \frac{W_0^2k_c}{2\pi}$ and where we set the disorder cut-off to be $k_c = \frac{\pi}{2a}$. To extract the localization length $\xi$ of the different eigenstates, we compute the corresponding inverse participation ratio $\text{IPR} = \int dx |\psi(x)|^4$ and average it over $1000$ realizations of the disorder. For weak disorder, the IPR is related to the inverse of the localization length via $\text{IPR} = \frac{1}{2\xi}$. The results for $\kF = k_c/2$ are shown in Fig.~\ref{fig:localization_length} for BSD, RSD and GCD. 
More details on the generation of each type of disorder, the average over the disorder realizations, as well as numerical results for other values of $k_c$ are given in the SM~\cite{supp_mat}. In particular, for $k_c < 2\kF$, we find $\xi$ of the order of the system size, consistently with effective delocalization.

For $k_c = 2\kF$, the three types of disorder exhibit clear differences showing the role of non-gaussianity and the presence of higher moments for the BSD and RSD. The difference between BSD and RSD may be attributed to odd-order disorder terms, which affect localization with opposite contributions~\cite{Lugan_Miniatura_2009_Anderson_localization_speckle,Piraud_Sanchez_Palencia_2011_Localization_wave_packet}, while the GCD has no such terms. In all cases, clear deviations from the usual $1/D$ white-noise behavior are found, showing the role of the vanishing spectral weight of the disorder at $k_c = 2\kF$. As discussed above, higher-order terms in the disorder are expected to give a non-vanishing spectral weight of the disorder at $2\kF$, and thus to a large but finite localization length. However, such terms are expected to give at least a $1/D^2$ behavior (note that third-order terms in SD have the same threshold at $k_c$ as the second-order term). None of the three disorders is apparently compatible with such a scaling. For GCD and RSD, a scaling compatible with $D^{-3/2}$ is visible in the weak disorder range and before the localization length is limited by finite-size effects. The precise dependence for BSD is more difficult to ascertain owing to the limited range of the data, but seems to show a crossover towards also $D^{-3/2}$ at weak disorder. The origin of such a $D^{-3/2}$ scaling is not understood at the moment and is clearly an important target for more numerical investigations and further studies. 

In summary we have studied the localization effects for interacting one-dimensional quantum particles in the presence of a colored random potential with linear vanishing of its Fourier components at the backscattering wavevector $2\kF$. Such a disorder is directly inspired, and generalizes the correlations of a speckle disorder. We showed that such a disorder leads to a new critical point for the localization-delocalization transition between a TLL and a phase in which the disorder is relevant. The properties of such a phase, and in particular its transport properties are yet a challenge. Indeed we examined the behavior of localization length on the non-interacting critical line and found an unexpected potential scaling $1/D^{3/2}$ as a function of the strength of the disorder $D$. Understanding and exploring the properties of such colored disorder is thus a strong challenge. On the theoretical side further analysis for the non-interacting case are clearly needed. For the interacting case, although on much shorter systems, tensor network or Monte-Carlo calculations should also allow an exploration of the phase diagram and of the shift of the transition point. On the experimental side true speckle potentials as well as potentials created by DMD \cite{Rubio_Abadal_Gross_2019_MBL_cold_atoms_DMD} (which could implement both GCD or SD) should allow to probe the effects studied in this paper. The need to match the cut-off of the disorder spectrum $k_c$ to $2\kF$ (or $2\pi\rho_0$ for bosons) would require to perform 
such experiments in a box potential rather than in the presence of a parabolic trap.

\begin{acknowledgments}
This work was supported
by the Swiss National Science Foundation under Division II (Grant No. 200020-219400),
the Agence Nationale de la Recherche (ANR, project ANR-CMAQ-002 France~2030),
and GENCI-TGCC (Grant 2023-A0110510300).
\end{acknowledgments}

\end{document}


\title{
Supplementary material for \\``Localization Transition for Interacting Quantum Particles in Colored-Noise Disorder''}

\author{Giacomo Morpurgo}
\affiliation{Department of Quantum Matter Physics, University of Geneva, 1205 Geneva, Switzerland}
\author{Laurent Sanchez-Palencia}
\affiliation{CPHT, CNRS, Ecole Polytechnique, IP Paris, F-91128 Palaiseau, France}
\author{Thierry Giamarchi}
\affiliation{Department of Quantum Matter Physics, University of Geneva, 1205 Geneva, Switzerland}

\date{\today}

\maketitle

\appendix
\section{Uncorrelated Gaussian disorder, speckle disorder, and Gaussian Coloured disorder}
\label{app:Disorder}

As a reminder and to ease the comparison between the different disorder distributions which are mentioned in the main text, we detail them a bit more in this section. 

\paragraph{Uncorrelated Gaussian disorder}
The disorder distribution which is used as a comparison benchmark through all of the letter is an uncorrelated disorder arising from a Gaussian distribution. A given realization of the disorder $V$ is then drawn from the following probability distribution :
\begin{equation}
    P(V) = e^{-\frac{1}{2D}\int dx V(x)^2} = e^{-\frac{1}{2D\Omega}\sum dq V_qV^*_q}
\end{equation}
Its correlations are then given by 
\begin{align}
    \overline{V(x)V(y)} &= D\delta(x-y)\\
    \overline{V_kV_{-k'}} &= D\Omega\delta_{k,k'}
\end{align}

\paragraph{Speckle disorder}
On the other hand, speckle disorder (SD), which is the main focus of this letter is instead correlated spatially, with correlation functions:

\begin{align}
\label{eq:app_correlation_space}
    \overline{W(x)W(y)} &= \frac{W_0^2k_c}{2\pi} \left(1 + \frac{\sin^2\left((x-y)\frac{k_c}{2}\right)}{\left((x-y)\frac{k_c}{2}\right)^2}\right)\\
    \overline{W_kW_{-k'}} &= W_0^2\left(1 - \frac{|k|}{k_c}\right)\theta(k_c-|k|)\,\Omega\,\delta_{k,k'}
\end{align}
in real and momentum spaces, respectively.

As can be seen in Eq.~\eqref{eq:app_correlation_space}, the correlations of this disorder are translation invariant in real space. This leads to the nice property of the correlations that they are diagonal in $k$-space, but with a coefficient depending on $k$ and which is $0$ above a certain $k_c$. This disorder potential originates from a complex random electric field $\mathcal{E}$. Both real and imaginary parts of its Fourier components $\mathcal{E}_k$ for $|k|<\frac{k_c}{2}$ are random variables originating from a Gaussian distribution $P(\text{Re/Im}(\mathcal{E}_k))=\frac{1}{\sqrt{2\pi\sigma}}e^{-\frac{1}{2}\frac{\text{Re/Im}(\mathcal{E}_k)^2}{\sigma^2}}$ with variance $\sigma = \left(\frac{\Omega W_0}{|C|}\right)^{1/2}\left(\frac{\pi}{2k_c}\right)^{1/4}$. The factor $\sqrt{\Omega}$ comes when we consider $P(\mathcal{E})$ in its entirety in Fourier space. The potential $W(x)$ felt by the atoms is then given by : 
\begin{equation}
\label{eq:app_SD_potential_space}
    W(x) = C\overline{\langle|\mathcal{E}|^2\rangle}\left(\frac{|\mathcal{E}(x)|^2}{\overline{\langle|\mathcal{E}|^2\rangle}}-1\right),
\end{equation}
where,  $C$ is a constant which depends on experimental parameters. We take it to be $>0$ for the blue-detuned disorder (BSD) and $<0$ for the red detuned one (RSD). The quantity $\overline{\langle|\mathcal{E}|^2\rangle} = \frac{\sigma^2k_c}{\Omega \pi}$ is the spatial and disorder average of $|\mathcal{E}|^2$. The spatial correlation of $W(x)$ obey then \eqref{eq:app_correlation_space}.

\paragraph{Gaussian colored disorder (Speckle-like disorder)} There exists a Gaussian correlated disorder which has the same second order correlations as the speckle disorder. A given realization of this disorder $W_{\text{GCD}}$ is given by the following probability distribution 

\begin{align}
    \label{eq:app_probability_distribution}
    P(W_{\text{GCD}}) &= e^{\frac{1}{2}\int dx \int dy \,\widetilde{A}^{-1}(x-y)W_{\text{GCD}}(x)W_{\text{GCD}}(y)}\\
    &=e^{-\frac{1}{2\Omega}\sum_{|k|<k_c} A_k^{-1}W_{\text{GCD},k}^*W_{\text{GCD},k}}\\
    A_k &= W_0^2\left(1-\frac{|k|}{k_c}\right)\\
    \widetilde{A}(x) &= \frac{1}{\Omega}\sum_k \, e^{ikx}A_k
\end{align}
Since there is no distinction between the SD and the GCD at the two-point correlation level, all the calculations which involve only two-point correlation will be identical for both.

In all cases, the average over the realizations of the disorder of an observable $A$ is then the result of the following expression :
\begin{equation}
    \overline{A} = \frac{\int d P(W) A(W)}{\int d P(W)}
\end{equation}

\section{Forward disorder}
In the derivation of the RG equations of the system, we neglected the forward disorder, we now comment more on the effects of forward disorder. For one dimensional systems, the effects of forward and backward disorder are completely decoupled, and we can consider them as being two independent realizations of the disorder : $W_b$ and $W_f$. The contribution of $W_f$ can be absorbed in the backward disorder by the following shift of the field $\phi$.
\begin{equation}
    \widetilde{\phi} = \phi -\frac{K}{u}\int^x dy \, W_f(y)
\end{equation}
This shift of $\phi$ leads to a change in the correlations for the new disorder $\widetilde{W_b}$ since it contains $\phi$ :
\begin{equation}
    \overline{\widetilde{W_b}(x)\widetilde{W_b}^*(y)} = \overline{W_b(x)W_b(y)}\overline{e^{i\frac{2K}{u}\int_y^xdy\,W_f(y)}}
\end{equation}
After performing the disorder average over $W_f$ and performing the spatial integrals, we get : 
\begin{equation}
    \overline{\widetilde{W_b}(x)\widetilde{W_b}^*(y)} = \overline{W_b(x)W_b(y)}e^{-2\frac{K^2}{u^2}\frac{W_0^2k_c}{2\pi}f(x-y)}
\end{equation}
with
\begin{multline}
 f(x-y) = \frac{2}{k_c^2}\left[-2 -2\gamma +2\cos\left[(x-y)k_c\right]\right. \\
  \quad + (x-y)k_c\textrm{Si}[(x-y)k_c] \\
 \quad \left.-2\log(|x-y|k_c)+2\textrm{Ci}(|x-y|k_c)\right]
\end{multline}
where $\gamma$ is the Euler constant, Si is the sine integral and Ci the cosine integral. $f(x)$ is an even function which for large $x$ behaves linearly as : $f(x)\approx \frac{\pi}{k_c}|x-y| $. The backward correlations can then be written as :
\begin{equation}
    \overline{\widetilde{W_b}(x)\widetilde{W_b}^*(y)} = \overline{W_b(x)W_b(y)}e^{-\frac{|x-y|}{\xi_f}}
\end{equation}
with a characteristic forward disorder length 
\begin{equation}
    \xi_f = \frac{u^2}{K^2}\frac{1}{W_0^2}
\end{equation}
which is reminiscent of the one for the uncorrelated Gaussian disorder $\xi_f = \frac{u^2}{2K^2D}$. \cite{Book_Giamarchi_2004_1D}.

The backward correlations are then exponentially suppressed with this length $\xi_f$. If this length is large enough, there is no effect on the correlations at order $W_0^2$ in the disorder. This is the case in particular for weak disorders, where $W_0^2$ is small. In this regime, neglecting the forward disorder is completely justified. At higher orders, the forward scattering will modify the correlations of the backward terms, leading to a small Fourier component at $2\kF$ of order $W_0^4$. 

\section{Derivation of the RG equations in the bosonization formalism}
For this derivation, we follow the procedure of \cite{Vidal_Giamarchi_2001_quasiperiodic_potential_long}. We can express the disorder by its Fourier components :
\begin{equation}
    W(x) =\frac{1}{\Omega}\sum_q W_q e^{iqx}
\end{equation}
where $\Omega$ is the volume of the system. The part of the action responsible for the disorder scatterings is then given by :
\begin{multline}
    S_w=\frac{1}{2\pi u\alpha}\int dx\int_0^\beta d\tau \sum_q \frac{W_q}{\Omega} \\ \times\left[e^{i(q^+x-2\phi(x,\tau))} + e^{i(q^-x-2\phi(x,\tau))}\right]
\end{multline}
where $q^\pm = q \pm2\kF$, $\tau$ the imaginary time, $\beta = \frac{1}{k_BT}$ with $k_B$ Boltzmann constant and $T$ the temperature of the system, which we put to $0$. The quantity which we are computing is :
\begin{multline}
    \langle T_\tau e^{i\sqrt{2}\phi(x',\tau')}  e^{-i\sqrt{2}\phi(x,\tau)}\rangle \\
    = \frac{\int\mathcal{D}\phi e^{i\sqrt{2}\phi(x',\tau')}  e^{-i\sqrt{2}\phi(x,\tau)}e^{-S_0-S_W}}{\int\mathcal{D}\phi e^{-S_0-S_W}}
\end{multline}

Performing an expansion in powers of $W_q$ up to order $2$, we get the following terms: 
\begin{align}
\langle T_\tau e^{i\sqrt{2}\phi(x',\tau')}  e^{-i\sqrt{2}\phi(x,\tau)}\rangle = I_0 - \frac{1}{2\pi u\alpha}I_1 +  \frac{1}{8(\pi u\alpha)^2} I_2
\end{align}

The ensemble averages that we will have to do are now with the action without disorder $S_0$. They are therefore of the shape \cite{Book_Giamarchi_2004_1D} :
\begin{multline}
    \label{eq:app_correlation_general_1D}
    \langle T_\tau e^{ic_1\phi(x_1,\tau_1)} \dots e^{ic_n\phi(x_n,\tau_n)}\rangle_0 \\
    \approx 
    \begin{cases}
        e^{-\frac{K}{2}\sum_{i>j}c_ic_j\ln\left(\frac{|r_i-r_j|}{\alpha}\right)}  &\sum_i c_i = 0\\
        0   &\sum_i c_i \neq 0
    \end{cases}
\end{multline}
where $ = (x,u\tau)$ and $r^2 = x^2 + u^2\tau^2$ and in the limit where $|r_i-r_j|\gg \alpha$. We will also use the following notation $F(r_i-r_j) = K\ln(|r_i-r_j|/\alpha)$.

With all of this, $I_0$ is therefore given by  :
\begin{equation}
    I_0 = \langle T_\tau e^{i\sqrt{2}\phi(x',\tau')}  e^{-i\sqrt{2}\phi(x,\tau)}\rangle_0 = e^{-F(r-r')} 
\end{equation}
$I_1$ is equal to 0 because of (\ref{eq:app_correlation_general_1D}). The interesting term is $I_2$, which will yield our RG equations. It reads
\begin{widetext}
\begin{multline}  
    I_2 = \frac{1}{\Omega^2}\sum_{q_1,q_2}W_{q_1}W_{q_2}\int d^2r_1 \int d^2r_2 \\
    \cdot \left\{e^{i(q_1^-x_1+q_2^+x_2)}\left[ \langle e^{i(\sqrt{2}\phi(r) -\sqrt{2}\phi(r') + 2\phi(r_1)-2\phi(r_2))}\rangle_0 - \langle e^{i(\sqrt{2}\phi(r) -\sqrt{2}\phi(r')}\rangle_0\langle e^{i(2\phi(r_1) -2\phi(r_2)}\rangle_0 \right] \right.\\
    + \left.e^{i(q_1^+x_1+q_2^-x_2)}\left[ \langle e^{i(\sqrt{2}\phi(r) -\sqrt{2}\phi(r') + 2\phi(r_2)-2\phi(r_1))}\rangle_0- \langle e^{i(\sqrt{2}\phi(r) -\sqrt{2}\phi(r')}\rangle_0\langle e^{i(2\phi(r_1) -2\phi(r_2)}\rangle_0 \right]\right\}
\end{multline}
\end{widetext}
Evaluating the ensemble average on $S_0$, doing the average on disorder realizations, reorganizing the terms, and doing the following change of coordinates : $R = (r_1+r_2)/2, \widetilde{r} = r_1- r_2$, we have the following expression for $I_2$.
\begin{widetext}
\begin{equation}
    I_2 = \frac{W_0^2}{\Omega}\sum_q\left(1-\frac{|q|}{k_c}\right)\theta(k_c+q)\theta(k_c-q)
     \int d^2R\int d^2\widetilde{r}\,e^{-2F(r-r')}e^{-2F(\widetilde{r})}2\cos(q\widetilde{x})e^{-i2\kF\widetilde{x}}\left[e^{\sqrt{2}\boldsymbol{\widetilde{r}}\cdot\boldsymbol{\nabla_Z}\left.F(z)\right|^{r'-R}_{r-R}}-1\right]
\end{equation}
\end{widetext}
Expanding the exponential in the square brackets, the zeroth order cancels out, the first order is equal to $0$ by symmetry after the sum on $q$. The second order term leads to terms in $\nabla_X^2 - \nabla^2_Y$ and $\nabla_X^2 + \nabla^2_Y$. The crossed terms $\nabla_X\nabla_Y$ are $0$ by parity in $\widetilde{y}$. The first of the two surving terms renormalizes the speed $u$. However, since the correction will be of order $W_0^2$, they can be neglected in the RG equations which we will find. Using $\int d^2R\left[F(r-R)-F(r'-R)\right](\nabla_X^2 + \nabla^2_Y)\left[F(r-R)-F(r'-R)\right] = -4\pi K F(r-r')$, passing in polar coordinates for $R$ and integrating on the angle $\theta$ (note that $\theta$ goes only from $0$ to $\pi$, since $u\tau$ is only positive), the second term becomes :
\begin{widetext}
\begin{equation}
    I_2 = e^{-F(r-r')}KF(r-r')\int d\widetilde{r} \,\widetilde{r}^3e^{-2F({r})}\frac{1}{\Omega}\sum_q 4\pi^2 W_0^2\left(1-\frac{|q|}{k_c}\right)\theta(k_c+q)\theta(k_c-q)\left[J_0((q+2\kF)\widetilde{r}) + J_0((q-2\kF)\widetilde{r})\right]
\end{equation}
\end{widetext}

Combining $I_0$ and $I_2$, after reexponentiation of the term in $W_0^2$, we can define an effective Luttinger parameter $K^{\text{eff}}$ we obtain using the notation $y = \frac{\alpha W_0}{u}$ as defined in the main text.
\begin{widetext}
\begin{equation}
    K^{\text{eff}} = K -\frac{K^2y^2}{2\alpha^4}\int_\alpha^{\infty}d\widetilde{r}\frac{1}{\Omega}\sum_q\left(1-\frac{|q|}{k_c}\right)\theta(k_c-|q|)\left[J_0((q+2\kF)\widetilde{r}) + J_0((q-2\kF)\widetilde{r})\right]
\end{equation}
\end{widetext}
To get the RG equations, we need to look at a small increase of the cutoff $\alpha(l+dl) = \alpha(l)e^{dl}$. $K^\text{eff}$ being a physical observable, it should not depend on the cutoff, which then leads to the RG equation (5) of the main paper. Replacing the Bessel functions by closed windows of size $1/\alpha(l)$ around $2\kF$ as described in the main text, we can perform the sum on the momenta and get the RG equations.

For the case $k_c>2\kF$, we have that the sum on the momenta yields in the continuum and for $\alpha(l)$ large enough such that $k_c > 2\kF + \frac{1}{\alpha(l)}$:
\begin{equation}
    \frac{1}{2\pi}\left(\int_{-2\kF-\frac{1}{\alpha(l)}}^{-2\kF+\frac{1}{\alpha(l)}} +\int_{2\kF-\frac{1}{\alpha(l)}}^{2\kF+\frac{1}{\alpha(l)}}\right) dq \left(1-\frac{|q|}{k_c}\right)= \frac{4(k_c-2\kF)}{2\pi\alpha(l)}
\end{equation}
We then can extract the RG equation by varying the cutoff $\alpha(l)$.
\begin{align}
     \frac{\partial K}{\partial l}& = -\frac{K^2(k_c-2\kF)}{\pi k_c}\frac{y^2}{ \alpha(l)} \\
     \frac{\partial y}{\partial l}& = (2 - K )y 
\end{align}
By absorbing a factor $\sqrt{\alpha(l)}$ in the definition of $y = \sqrt{\alpha(l)}\tilde{y}$, these equations can be recasted in : 
\begin{align}
     \frac{\partial K}{\partial l}& = -\frac{K^2(k_c-2\kF)}{\pi k_c}\tilde{y}^2 \\
     \frac{\partial \tilde{y}}{\partial l}& = (\frac{3}{2} - K )\tilde{y} 
\end{align}
where it is quite clear that we recover, as in the case of uncorrelated Gaussian disorder, that the transition point is at $K =3/2$.

For the case $k_c=2\kF$, the sum on the momenta yields a different result :
\begin{equation}
    \frac{1}{2\pi}\left(\int_{-k_c}^{-k_c+\frac{1}{\alpha(l)}} +\int_{k_c-\frac{1}{\alpha(l)}}^{k_c}\right) dq \left(1-\frac{|q|}{k_c}\right)= \frac{1}{2\pi\alpha^2(l)}
\end{equation}

We then can extract the RG equation by varying the cutoff $\alpha(l)$
\begin{align}
     \frac{\partial K}{\partial l}& = -\frac{K^2}{4\pi k_c}\frac{y^2}{ \alpha^2(l)} \\
     \frac{\partial y}{\partial l}& = (2 - K )y 
\end{align}
By absorbing a factor $\alpha(l)$ in the definition of $y = \alpha(l)\tilde{y}$, these equations can be recasted in : 
\begin{align}
     \frac{\partial K}{\partial l}& = -\frac{K^2}{4\pi k_c}\tilde{y}^2 \\
     \frac{\partial \tilde{y}}{\partial l}& = (1 - K )\tilde{y}
\end{align}
The advantage of this writing is that again there is no $\alpha(l)$ in the equations. From these expressions, it naturally follows that the transition point occurs at $K = 1$.

\section{Derivation of the RG equations for interacting fermions perturbative in interactions}
To get the renormalizations equations perturbatively in the interaction $g$ and disorder $W_0^2$, we have to compute the following correlation functions, which correspond respectively to the backward scattering and interaction diagram.

For the bacward scattering, we have to compute the following expression
\begin{multline}
    \label{eq:app_fermions_correlation_disorder}
   \Big \langle T_\tau\left[c_{-r}(k+q,\tau_4)c_r(k'-q,\tau_3)\right.\\
    \left.c^\dagger_{-r}(k', \tau_2)c^\dagger_{r}(k,\tau_1)e^{-\int_0^\beta d\tau'H_{g,W_0}} \right] \Big\rangle_{H_0}
\end{multline}
As in the main text, $r$ denotes the right and left movers, $\varepsilon_r=\pm 1$ correspondingly. Here $H_0$ is just the free Hamiltonian without disorder or interaction, and $H_{g,W_0}$ contains the interaction and backward scattering terms of equation (9) of the main text. 

Instead, for the interaction, we have to compute the following expression~:
\begin{multline}
    \label{eq:app_fermions_correlation_interaction}
    \Big\langle T_\tau\left[c_r(k+q,\tau_4)c_{-r}(k'-q,\tau_3)\right.\\
     \left. c^\dagger_{-r}(k',\tau_2)c^\dagger_{r}(k,\tau_1)e^{-\int_0^\beta d\tau'H_{g,W_0}}\right] \Big\rangle_{H_0}
\end{multline}

\begin{figure}
\includegraphics[width=0.9\columnwidth]{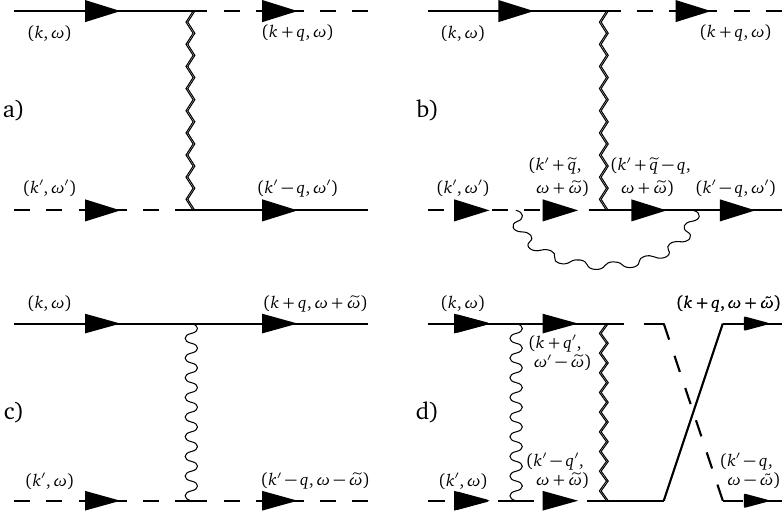}
\caption{\label{fig:diagrams} 
Diagrams contributing to the renormalization of interactions and disorder up to order two in $g$ and $W_0^2$. The solid (resp.~dashed) lines denote right (resp.~left) movers. The wiggly line is the interaction $g$, while the zig-zag line indicate the averaged disorder line. The frequencies written are Matsubara frequencies. The panels a) and b) show,  respectively, first and second order disorder backscattering diagrams. Frequencies are conserved separately on the top and bottom lines for these diagrams since the disorder is time independent. The panels c) and d), respectively, show first and second orders of the interaction diagram. In such a diagram, frequency can be exchanged between the top and bottom lines.  
}
\end{figure}

We perform an expansion of the exponential up to 2nd order in $g$, $W_0^2$ and apply Wick's theorem to the creation/annihilation operators since $H_0$ is quadratic. Each pair of creation/annihilation can be written as a free electron Green's function/propagator in imaginary time:
\begin{align}
    G_{o,r}(k,\tau) &= -\langle T_\tau c_r(k,\tau)c^\dagger_r(k,0)\rangle
\end{align}
The Green's function admits a Matsubara frequency representation $G(k,\tau) = \frac{1}{\beta}\sum_{i\omega_n}G(k,i\omega_n)e^{-i\omega_n\tau}$, with $\omega_n = (2n+1)\pi/\beta$  \cite{Book_Mahan_2000_many_body_physics}. For simplicity of notation, we drop the index $n$ from $\omega_n$ in the following. The Matsubara free electron propagator has the shape :
\begin{align}
    G_{o,r}(k,\omega) &= \frac{1}{i\omega-v_\text{F}(\epsilon_rk-\kF)}
\end{align}

The first order term of \eqref{eq:app_fermions_correlation_disorder}  can now be related to the diagram shown in Figure~\ref{fig:diagrams}.a) , while the first order term of Eq.~\eqref{eq:app_fermions_correlation_interaction} can now be related to the diagram shown in Figure~\ref{fig:diagrams}.c).  Each line is the free electron propagator (continuous lines are right movers and dashed lines are left movers). The zig-zag line denotes the backscattering process, which carries a momentum $q$ but no frequency and has a weight given by $\overline{W_qW_{-q}} = A(q)\theta(k_c-|q|)$. The wiggly line denotes the interaction process, which carries a momentum $q$, a frequency $\widetilde{\omega}$ and has a weight of $g$.

The second order in the expansion of \eqref{eq:app_fermions_correlation_disorder} leads in particular to the diagram Figure~\ref{fig:diagrams}.b), which renormalizes the diagram Figure~\ref{fig:diagrams}.a), while for \eqref{eq:app_fermions_correlation_interaction} we find the diagram Figure~\ref{fig:diagrams}.d) which renormalizes Figure~\ref{fig:diagrams}.c). We then have to perform the integrals/sums over the internal degrees of freedom \cite{Book_Mahan_2000_many_body_physics} of Figure~\ref{fig:diagrams}.b) (and Figure~\ref{fig:diagrams}.d)) to find the renormalization of $A(q)\theta(k_c-|q|)$ (and $g$). For the diagram in Figure~\ref{fig:diagrams}.b) the variables on the external legs can expressed as~: $k = \kF + \delta k$, with $\delta k$ small since this leg corresponds to a right mover and $k' = -\kF - \delta k'$ with $\delta k'$ small, since this leg corresponds to a left mover. The result is independent of the external frequencies and diverges logarithmically in the cutoff on the momenta $\Lambda$ as~: $-\frac{gA(q)\theta(k_c-|q|)}{4\pi v_\text{F}}\ln\left[\frac{4\Lambda^2}{(q-2\kF)^2}\right]$. For the diagram in Fig.~\ref{fig:diagrams}.d) which renormalizes the interaction term $g$, we make the same choice of $k$ and $k'$ as before. To extract the main divergence, there is a smart choice of the external frequencies which simplifies the expressions \cite{Review_Solyom_1979_Fermi_gas_1d_conductor}. We thus take $\omega = -\frac{1}{2} \widetilde{\omega}$ and $\omega' = \frac{3}{2}\widetilde{\omega}$. We find that the leading divergence for $q = 0$ is also logarithmic in the cutoff and is given by : $-\frac{gW_0^2}{4\pi k_c v_\text{F}^2}\ln\left(\frac{v_\text{F}^2\Lambda^2}{\widetilde{\omega}^2/4}\right)$.

There are two more diagrams of order $g^2$, which could be relevant in the interaction renormalization. The first has two parallel interaction lines and the second has crossed interaction lines. However, their divergences in $\Lambda$ cancel exactly and they do not intervene in our RG. We therefore find an effective interaction $g^{\text{eff}}$ and an effective disorder strength ${W_0^{2}}^\text{eff}$ given by :

\begin{align}
    g^{\text{eff}} &= g -\frac{gW_0^2}{4\pi k_cv_\text{F}^2}\ln\left(\frac{4v_\text{F}^2\Lambda^2}{\widetilde{\omega}^2}\right)\\
    {W_0^{2}}^\text{eff}\theta(k_c-|q|) &= W_0^{2}\theta(k_c-|q|) \\
    &\quad \quad -\frac{gW_0^2}{4\pi k_cv_\text{F}^2}\ln\left[\frac{4\Lambda^2}{(q-2\kF)^2}\right]\theta(k_c-|q|)\nonumber
\end{align}

The final step to get the RG equations is then to keep the effective variables constant when we decrease the cutoff $\Lambda(l+dl)=\Lambda(l)e^{-dl}$. We then obtain the RG equations (10) and (11) of the main text.

\section{Numerical generation of the disorder}
To find the localization length of the system, we need to generate the disorder following different disorder distributions.  For the Gaussian colored disorder, since the disorder correlations are translation invariant, we generate first each Fourier component of the disorder $W_q$ with the probability distribution given by \eqref{eq:app_probability_distribution}. Once we have all of the $W_q$ we can Fourier transform them back to have the disorder $W(x)$ in real space. Note that $W(x)$ is real, therefore there is a condition on its Fourier component $W_q = W_{-q}^*$.

For the real speckle disorder (both blue and red), we start by generating the complex Fourier components of the electric field $\mathcal{E}_q$ as defined in section~\ref{app:Disorder}. We then Fourier transform it to have the electric field in space and compute $W(x)$ using \eqref{eq:app_SD_potential_space}. We took the numerical values $C=1$, which generates a BSD. Because of the particle-hole ($d^\dagger_i = (-1)^ic_i$) symmetry of the Hamiltonian when also changing the sign of $W_i$, we could find the RSD properties by looking at $k_\text{F,RSD} = \pi -k_\text{F,BSD}$. 

\section{Average over the disorder of the IPR}
For each Hamiltonian generated, we compute the IPR as a function of the energy. Since we have a finite number of sites ($10000$), we have a finite number of eigenvalues of the energy. To perform the average over the disorder, we have to bin our energy space. We do this by selecting the minimal and maximal eigen energies among all $1000$ disorder realizations. We then separate this interval in $400$ bins of same size $\Delta E$. For each realization $i$ of disorder, we compute the IPR associated to a given bin $\text{IPR}^{i}_{E,E+\Delta E}$ by averaging over all eigenstates which fall in the corresponding energy range $\text{IPR}^i_{E,E+\Delta E}=1/N^i_{E,E+\Delta E}\sum_{E<E_n<E+\Delta E} \text{IPR}^i(E_n)$. Finally, we average this IPR for the same bin over all realizations $N_\text{real}$~: $\text{IPR}_{E,E+\Delta E}=\frac{1}{N_\text{real}}\sum_i\text{IPR}^i_{E,E+\Delta E}$.\\

\section{Localization length when $k_c<2\kF$}
For $k_c < 2\kF$, we find that $\xi$ is of the order of the system size, which suggests that there is no localization. This is consistent with the results of \cite{Lugan_Miniatura_2009_Anderson_localization_speckle}, which showed that the second order Lyapunov exponent vanishes and only higher order can give a finite localization length. Such a length is too large for our numerical calculation to capture it. This shows clearly that although higher order harmonics of the disorder formally still lead to localization,  the system is practically not localized in that case. 

\section{Localization length when $k_c>2\kF$}
\begin{figure}[h]
\includegraphics[width=0.95\columnwidth]{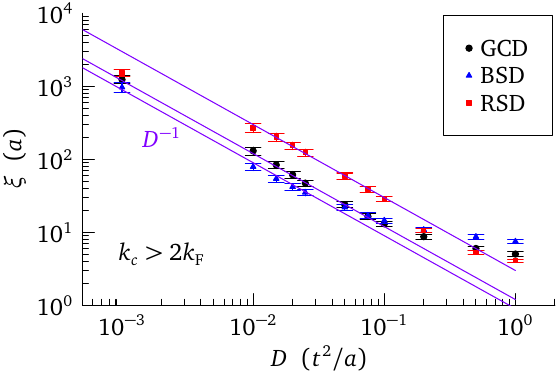}
\caption{\label{fig:localization_length_kf_3_8_kc} 
Localization length $\xi$ as a function of disorder strength $D$ for eigenstates at momentum $\kF = 3k_c/8$ of a noninteracting system with GCG (black circles), BSD (blue triangles) and RSD (red squares). The error bars indicate the standard deviation coming from the different disorder realizations.
The localization length shows the usual scaling $D^{-1}$ of the uncorrelated disorder. The lines plotted are
guides to the eye. 
}
\end{figure}

Using the same procedure as in the main paper, we look at the localization length scaling of the strength of the disorder, for $\kF = 3k_c/8$, which is an example of the regime $k_c >2\kF$, where there are finite components of backscattering. As shown in Figure~\ref{fig:localization_length_kf_3_8_kc}, we recover the $D^{-1}$ scaling associated to the uncorrelated Gaussian disorder, and this for the CGD, BSD and RSD. We also see here that we reach the strong disorder regime where the localization length does not follow the scaling anymore.

